\documentclass[10pt]{iopart}
\usepackage{epsfig}
\begin{document}

\title{QCD effects to Bjorken 
unpolarized sum rule for $\nu N$ deep-inelastic scattering} 

\author{S.I. Alekhin\dag\ and A.L. Kataev\ddag  
}

\address{\dag\ Institute of High Energy Physics, 
Protvino, Russia}

\address{\ddag\ Institute for Nuclear Research of the Academy 
of Sciences of Russia, 117312, Moscow, Russia} 
\ead{alekhin@sirius.ihep.su;~kataev@ms2.inr.ac.ru}

\begin{abstract}
The possibility of the first measurement 
of Bjorken unpolarized sum rule for $F_1$ structure function 
of $\nu N$ deep-inelastic scattering at Neutrino Factories is commented. 
The brief summary of various theoretical contributions
to this  sum rule is given. Using the next-to-leading set 
of parton distributions functions, we simulate the expected $Q^2$-behavior 
and emphasize that its measurement can allow 
to determine the value of the QCD strong coupling constant $\alpha_s$ 
with reasonable theoretical uncertainty, dominated by the 
ambiguity in the existing estimates of  
the twist-4 non-perturbative $1/Q^2$-effect.

\end{abstract}




The Bjorken unpolarized sum rule 
for $F_1$ structure function of $\nu N$ DIS, which was  theoretically
derived 
quite long ago in the classical work of Ref.~\cite{Bjorken:1968dy}, 
is still remaining experimentally unchecked.
However, at Neutrino Factories, due to large variation of 
$y=E_{had}/E_{\nu}\leq 1/(1+\frac{xM_W}{2E_{\nu}})$ it might  be possible 
to extract from the data for differential cross-sections data 
of $\nu N$ DIS  all 6 structure functions, including\
namely $xF_1^{\nu N}$, $xF_1^{\overline{\nu} N}$.

The direct measurement of  $F_1^{\nu N}$ SF is allowing to perform  
the first experimental determination of the Bjorken unpolarized sum rule, 
which has the following theoretical expression 
\begin{equation}
\label{Bj}
I_1(Q^2)=
 \int_0^{1}dx\bigg[F_1^{\nu n}(x,Q^2)-F_1^{\nu p}(x,Q^2)
\bigg]=1-\frac{2}{3}\frac{\alpha_s}{\pi}-...+O(\frac{1}{Q^2}) 
\end{equation}
The Bjorken unpolarized sum rule is related to the 
Adler isospin sum rule $I_2(Q^2)=\int_0^1(dx/2x)[F_2^{\nu n}(x,Q^2)-
F_2^{\nu p}(x,Q^2)]=1$ by the following relation $I_1(Q^2)=I_2+I_L(Q^2)$,
where $I_L(Q^2)=\int_0^1(dx/x)[F_L^{\nu p}(x,Q^2)-F_L^{\nu n}(x,Q^2)]$ 
is the Callan-Gross relation for $\nu N$ DIS, which contains 
perturbative QCD corrections 
and higher-twist terms, responsible for difference between 
the Bjorken unpolarized sum rule and the explicit Adler sum rule. 
In the $\overline{MS}$-scheme the massless  
perturbative QCD expression for $I_1$, 
namely $I_1^{PT}$,
is known analytically up to order $O(\alpha_s^3)$ corrections.
The leading order (LO), next-to-leading 
order (NLO) and next-to-next-to-leading order  corrections 
were subsequently  calculated in Refs.\cite{Altarelli:1978id},
\cite{Gorishnii:1983gs} and Ref.\cite{Larin:1990zw} respectively.
The numerical estimates of the order  $O(\alpha_s^4)$-contributions
are also available \cite{Kataev:1995vh,Samuel:1995jc,Broadhurst:2002bi}. 
Moreover, in Ref.\cite{Blumlein:1998sh}
the  heavy-quark mass dependence of 
$I_1$ was calculated  to the level of $\alpha_s^2$-terms.

However, there are also order $O(1/Q^2)$ contributions to $I_1$.  
Indeed, with taking into account the non-perturbative twist-4 contributions,
calculated within the context of operator product expansion, 
one can write down theoretical expression  
for  $I_1$ as
\begin{equation}
I_1(Q^2)=I_1^{PT}(Q^2)-\frac{8}{9}\frac{<<O>>}{Q^2} +O(\frac{1}{Q^4})
\end{equation}    
where 
$2p_{\mu}<<O>>=<p|O_{\mu}|p>$, 
$O_{\mu}=\overline{u}\tilde{G}_{\mu\nu}\gamma_{\nu}\gamma_{5}u-
\overline{d}\tilde{G}_{\mu\nu}\gamma_{\nu}\gamma_{5}d$
and  $\tilde{G}_{\mu\nu}=(\epsilon_{\mu\nu\alpha\beta}/2)G_{\mu\nu}^a
(\lambda^a/2)$ \cite{Shuryak:1981kj}. The numerical value of the twist-4 
term  was calculated in Ref.\cite{Braun:1986ty} with the help of 3-point 
functions QCD sum rules formalism. The result of this work is
$<<O>>=0.15\pm 0.07$ GeV$^2$, where we take for the theoretical error 
the conservative 
estimate 50$\%$.

\begin{figure}
\centerline{\epsfig{file=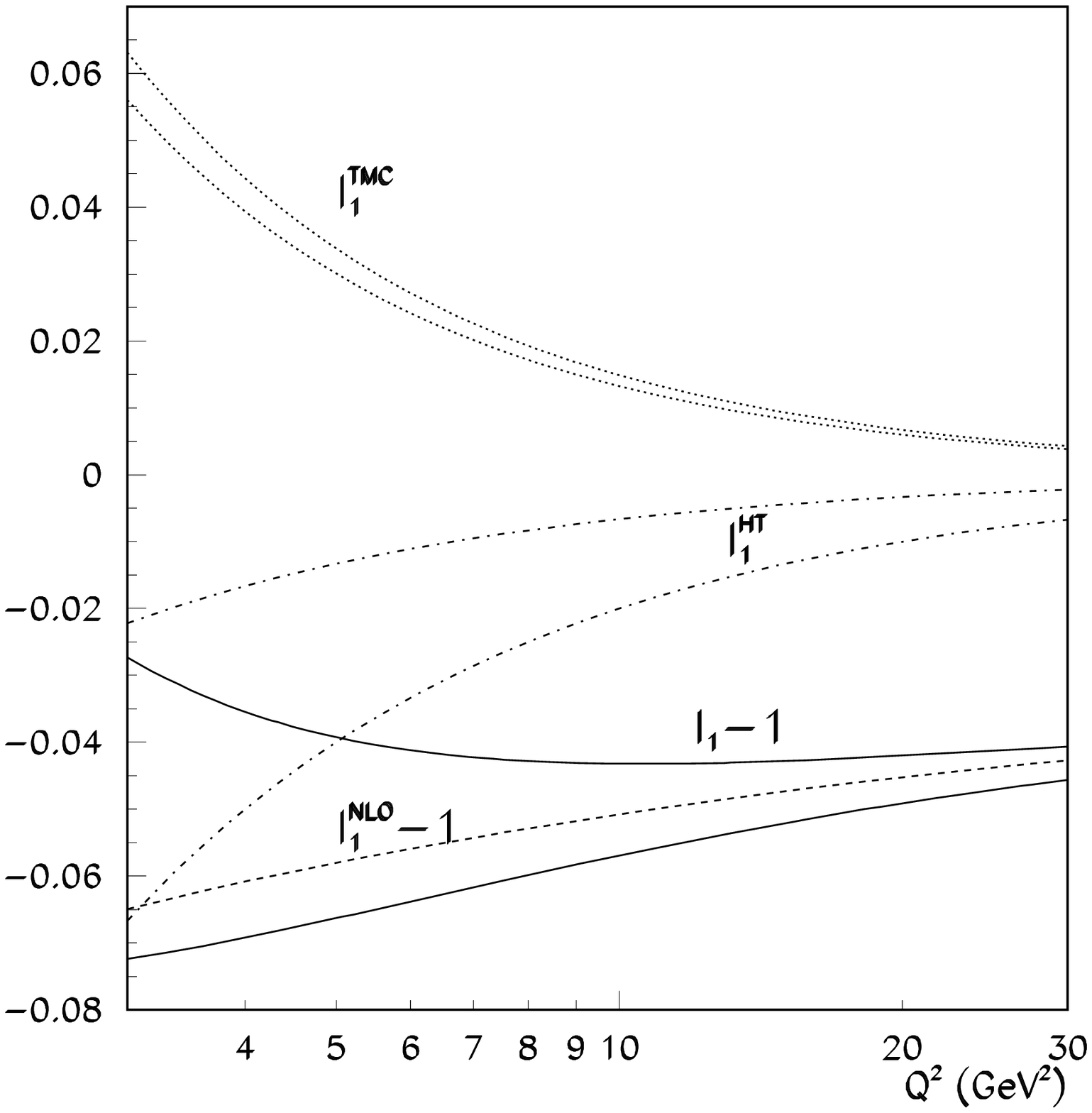,width=10cm,height=6.5cm}}
\caption{The perturbative contribution (dashes), 
the target mass correction (dots), and 
to the high-twist contribution (dashed dots) to $I_1(Q^2)$;
full curve give the sum of all these terms.}
\label{fig:bsr}
\end{figure}

\begin{figure}
\begin{center}
\epsfig{file=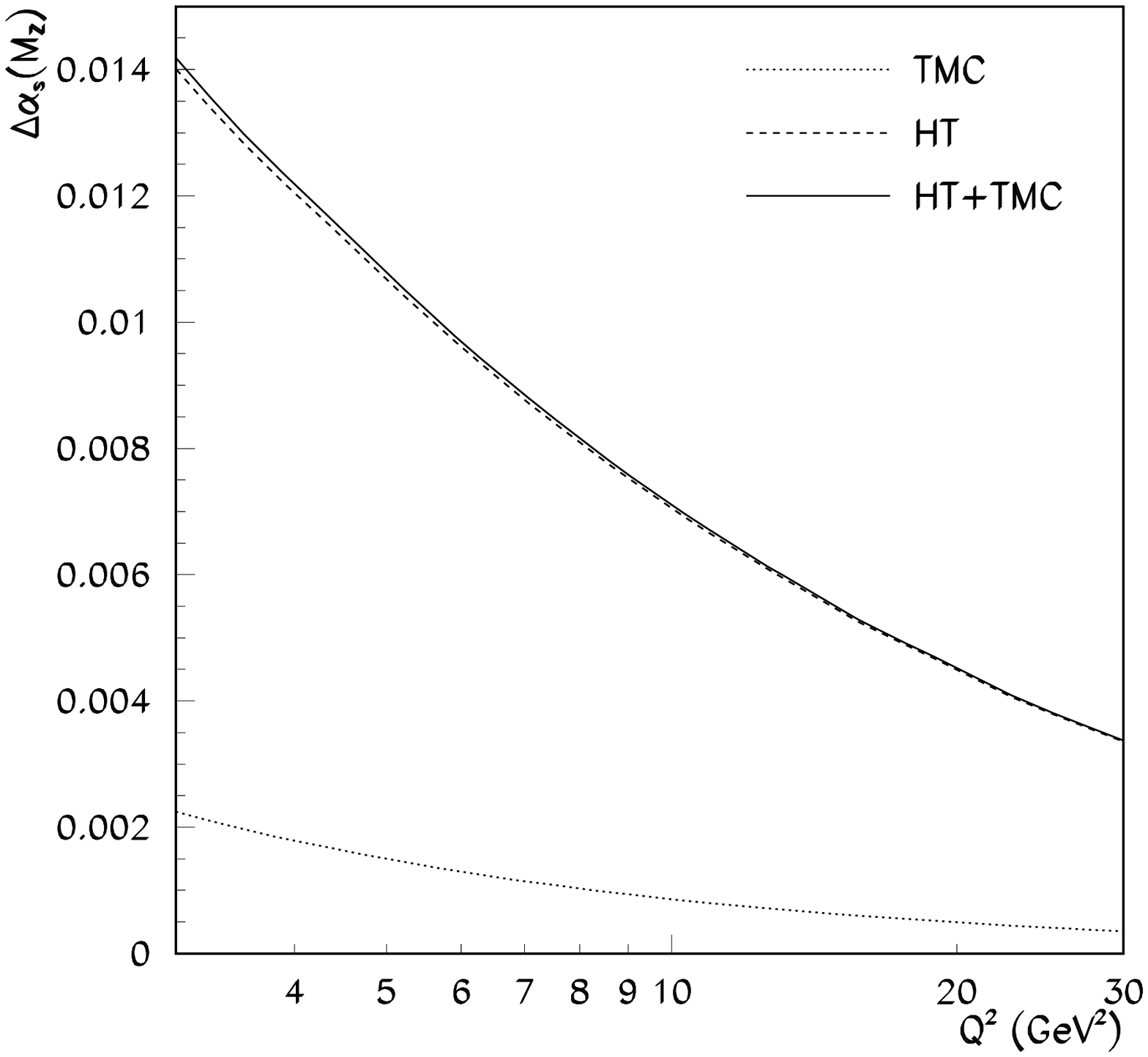,width=10cm,height=6.5cm}
\caption{The theoretical errors in $\alpha_{\rm s}$ extracted from the Bjorken 
sum rule (dashes: the error due to high-twist contribution; 
dots: the error due to target mass correction; 
full curve: combination of both).}
\end{center}
\label{fig:bsrerr}
\end{figure}

The $Q^2$-behavior of $I_1^{PT}$ in the NLO is given in Fig.\ref{fig:bsr} 
together with the high-twist contribution of Eq.(2) and the target 
mass (TMC)
correction term calculated using the parton distributions functions 
(PDFs) of Ref.\cite{Alekhin:2001ch}.
One can see that the main uncertainty in $I_1(Q^2)$ comes from 
the high-twist contribution, which rises at low $Q^2$. 
The theoretical error in the $\alpha_s(Q)$ value determined from the 
measurement of $I_1(Q^2)$ reads 
\begin{equation}
\Delta\alpha_s(M_Z)=\Delta I_1(Q^2)\bigg[\frac{dI_1(Q^2)}{d\alpha_s(Q^2)}
\frac{d\alpha_s(Q^2)}{d\alpha_s(M_Z)}\bigg]^{-1}, 
\end{equation}
where $\Delta I_1(Q^2)$ is the corresponding 
theoretical uncertainty in $I_1(Q^2)$.
The errors in $\alpha_s(Q^2)$ due to uncertainties in the high-twist 
term and the TMC are given in Fig.2  
(the latter is determined by the PDFs uncertainties).
At $Q^2=4$ GeV$^2$ and $Q^2=10$ GeV$^2$ the total theoretical error in 
$\alpha_s$ is 
\begin{equation}
\label{uns}
\Delta^{HT}\alpha_s(M_Z)=0.012~~~~~and~~~~ \Delta^{HT}\alpha_s(M_Z)=0.007
\end{equation}
correspondingly.

To conclude, the first experimental determination of the Bjorken unpolarized 
sum rule, which will be possible at Neutrino Factories, can be considered
as an additional source of precise determination of $\alpha_s$, provided 
the twist-4 contributions are known with less ambiguities. Moreover, turning 
the problems around and  fixing $\alpha_s(M_Z)$-value in the available 
perturbative expression for the Bjorken unpolarized sum rule,  
one can try to extract from the analysis of low $Q^2$-data of 
Neutrino Factories  the independent estimates of the corresponding 
twist-4 contributions. 
Another problem, which can be of interest for the future program of 
Neutrino Factories, is the study of nuclear corrections to this still 
unmeasured sum rule of $\nu N$ DIS (for the discussions see 
Ref.\cite{Mangano:2001mj}).

{\bf Acknowledgments} 
We are grateful to S.A. Kulagin for discussions.
The work was supported by RFBR Grant N 00-02-17432.
One of us (ALK) wishes to thank the OC of NuFact'02 Workshop for 
hospitality in London. His research was also supported by 
RFBR Grant N 02-01-00601.

\end{document}